\documentclass{elsart1p}
\usepackage{graphicx}

\usepackage{amssymb}
\begin{document}
\begin{frontmatter}
% notes:
% use the corauthref command within \author for corresponding author
% footnotes;
\title{Percolation and Deconfinement}

\author{Brijesh K Srivastava}

\address{Department of Physics, Purdue University, West Lafayette,
Indiana, USA}

\begin{abstract}
Possible phase transition of strongly interacting matter from hadron to a Quark-Gluon Plasma (QGP) state have in the past received considerable interest. It has been suggested that this problem might be treated by percolation theory. The Color String Percolation Model (CSPM) is used to determine the equation of state (EOS) of the QGP produced 
in central Au-Au collisions at RHIC energies. The bulk thermodynamic quantities- energy density, entropy density and the sound velocity- are obtained in the framework of CSPM. It is shown that the results are in excellent agreement with the recent lattice QCD calculations(LQCD).  

\end{abstract}

\end{frontmatter}

% main text
\section{Introduction}
\label{}
One of the main goal of the study of relativistic heavy ion collisions is to study the deconfined matter, known as Quark-Gluon Plasma(QGP), which is expected to form at large densities. It has been suggested that the transition from hadronic to QGP state can be treated by percolation theory \cite{celik}. The formulation of percolation problem is concerned with elementary geometrical objects placed on a random d-dimensional lattice. The objects have a well defined connectivity radius $\lambda$, and two objects can communicate if the distance between them is less than $\lambda$. Several objects can form a cluster of communication. At certain density of the objects a infinite cluster appears which spans the entire system. This is defined by the dimensionless percolation parameter $\xi$ \cite{isich}.  Percolation theory has been applied to several areas ranging from clustering in spin system to the formation of galaxies. Figure 1 shows the transition from disconnected to connected system at high densities.

\begin{figure}
\centering        
%\vspace*{-0.2cm}
\resizebox{0.70\textwidth}{!}{
\includegraphics{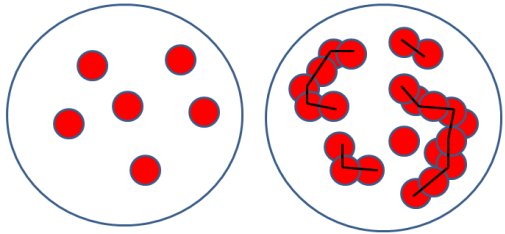}}
\vspace*{-1.0cm}
\caption{Left panel: Disconnected discs. Right panel: Overlapping discs forming a cluster of communication.} 
\label{perc1}
\end{figure}

 In nuclear collisions there is indeed, as a function of parton density, a sudden onset of large scale color connection. There is a critical density at which the elemental objects form one large cluster, loosing their independent existence. Percolation would correspond to the onset of color deconfinement and it may be a prerequisite for any subsequent QGP formation.

The determination of the EOS of hot, strongly interacting matter is one of the main challenges of strong interaction physics (HotQCD Collaboration) \cite{bazavov}. Recent LQCD calculations for the bulk thermodynamic observables, e.g. pressure, energy density, entropy density and for the sound velocity have been reported \cite{bazavov}. A percolation model coupled with hydrodynamics has been utilized to calculate these quantities from the STAR data at RHIC energies in central Au+Au collisions.
 
\section{Color String Percolation Model}   
Multiparticle production at high energies is currently described in terms of color strings stretched between the projectile and target. Hadronizing these strings produce the observed hadrons. The strings act as color sources of particles through the creation of $q \bar{q}$ pairs from the sea. The number of strings grows with the energy and with the number of nucleons of participating nuclei. Color strings may be viewed as small discs in the transverse space filled with the color field created by colliding partons. Particles are produced by the Schwinger mechanisms \cite{swinger1}.  With growing energy and size of the colliding nuclei the number of strings grow and start to overlap to form clusters \cite{pajares1,pajares2}. At a critical density a macroscopic cluster appears that marks the percolation phase transition. 2D percolation is a non-thermal second order phase transition,
but in CSPM the Schwinger barrier penetration mechanism for particle production and the fluctuations in the associated string tension due to the strong string interactions make it possible to define a temperature.
Consequently the particle spectrum is "born" with a thermal distribution \cite{pajares3,bialas}. The percolation threshold at which the spanning cluster appear, a "connected" system of color sources, identifies the percolation phase transition. 

With an increasing number of strings there is a progression from isolated individual strings to clusters and then to a large cluster which suddenly spans the area. In two dimensional percolation theory the relevant quantity is the dimensionless percolation density parameter given by \cite{pajares1,pajares2}  
\begin{equation}  
\xi = \frac {N S_{1}}{S_{N}}
\end{equation}
where N is the number of strings formed in the collisions and $S_{1}$
 is  the transverse area of the single string and $S_{N}$ is the transverse nuclear overlap area. The critical cluster which spans $S_{N}$, appears for
$\xi_{c} \ge$ 1.2 \cite{satz1}. As $\xi$ increases the fraction of $S_{N}$ covered by this spanning cluster increases.

The percolation theory governs the geometrical pattern of string clustering. It requires some dynamics to describe the interaction of several overlapping strings. It is assumed that a cluster behaves as a single string with a higher color field corresponding to the vectorial sum of the color charge of each individual string. Knowing the color charge, one can calculate the multiplicity $\mu_{n}$ and the mean transverse momentum $\langle p_{t}^{2} \rangle_{n}$ of the particles produced by a cluster of strings. One finds \cite{pajares1,pajares2}  
 
\begin{equation}  
\mu_{n}= \sqrt \frac {nS_{n}}{S_{1}}\mu_{1}
\end{equation}

\begin{equation}  
\langle p_{t}^{2} \rangle_{n}= \sqrt \frac {nS_{1}}{S_{n}}\langle p_{t}^{2} \rangle_{1}
\end{equation}
where $\mu_{1}$ and $\langle p_{t}^{2}\rangle_{1}$ are the mean multiplicity and average transverse momentum squared of particles produced by a single string with a transverse area
$S_{1} = \pi r^{2}_{0}$. In the saturation limit, all the strings overlap into a single cluster that approximately occupies the whole interaction area, one gets the following universal scaling law 

\begin{equation}                                      
\langle p_{t}^{2}\rangle_{n}  = \frac {S_{1}}{S_{n}}\frac {\langle p_{t}^{2}\rangle_{1}}{\mu_{1}}{\mu_{n}}    
\end{equation} 

 This scaling law shows a reasonable agreement with all the experimental data for all projectiles, targets and energies \cite{pajares4}. In the limit of high density one obtains
\begin{equation}
\langle \frac {nS_{1}}{S_{n}} \rangle = 1/F^{2}(\xi)
\end{equation}
with
\begin{equation}
F(\xi) = \sqrt \frac {1-e^{-\xi}}{\xi}
\end{equation}
being the color suppression factor. It follows that 
\begin{equation}
\mu = N F(\xi)\mu_{1},  \langle p_{t}^{2}\rangle = \frac{1}{F(\xi)}\langle p_{t}^{2}\rangle_{1}
\end{equation}
A similar scaling is found in the Color Glass Condensate approach (CGC)\cite{larry}. The saturation scale $Q_{s}$ in CGC corresponds to $ {\langle p_{t}^{2} \rangle_{1}}/F(\xi)$ in CSPM.  
The net effect due to $F(\xi)$ is the reduction in hadron multiplicity and increase in the average transverse momentum of particles. The CSPM model calculation for hadron multiplicities and momentum spectra was found to be in excellent agreement with experiment \cite{diasde,diasde2}. Within the framework of clustering of color sources, the elliptic flow, $\it v_{2}$, and the dependence of the nuclear modification factor on the azimuthal angle show reasonable agreement with the RHIC data \cite{bautista}. The critical density of percolation is related to the effective critical temperature and thus percolation may be the way to achieve deconfinement in the heavy ion collisions \cite{pajares3}. An additional important check of this interacting string approach was provided by the measurement of Long Range forward backward multiplicity Correlations (LRC) by the STAR group at RHIC \cite{LRC2}.

\section{Experimental Determination of the Percolation density Parameter $\xi$}

To obtain the value of $\xi$ from data, a parameterization of p-p events at 200 GeV  is used to compute the $p_{t}$ distribution \cite{nucleo}

\begin{equation}
dN_{c}/dp_{t}^{2} = a/(p_{0}+p_{t})^{\alpha}
\end{equation}
where a, $p_{0}$, and $\alpha$ are parameters used to fit the data. This parameterization  also can be used for nucleus-nucleus collisions to take into account the interactions of the strings\cite{nucleo}
\begin{equation}
p_{0} \rightarrow p_{0} \left(\frac {\langle nS_{1}/S_{n} \rangle_{Au-Au}}{\langle nS_{1}/S_{n} \rangle_{pp}}\right)^{1/4}
\end{equation}
where $S_{n}$ corresponds to the area occupied by the n overlapping strings.
The thermodynamic limit, i.e. letting n and $S_{n}$ $\rightarrow \infty$  while keeping $\xi$ fixed, is  used to evaluate $F(\xi)$ given by Eq.(5). For nucleus-nucleus collisions Eq.(8) becomes

\begin{equation}
dN_{c}/dp_{t}^{2} = \frac {a}{{(p_{0}{\sqrt {F(\xi_{pp})/F(\xi)}}+p_{t})}^{\alpha}}
\end{equation}
In pp collisions  $ \langle nS_{1} / S_{n} \rangle_{pp}$ $\sim$ 1  due to the low string overlap probability. 

The STAR analysis of charged hadrons had presented the preliminary results for the percolation density parameter, $\xi$ at RHIC for several collisions systems as a function of centrality\cite{nucleo}. Figure 2 shows $\xi$ as function of the number of participant nucleons($N_{part}$) in Au+Au collisions at $\sqrt{s_{NN}}=$ 200 and 62.4 GeV. 
\begin{figure}
\centering        
%\vspace*{-0.2cm}
\resizebox{0.50\textwidth}{!}{
\includegraphics{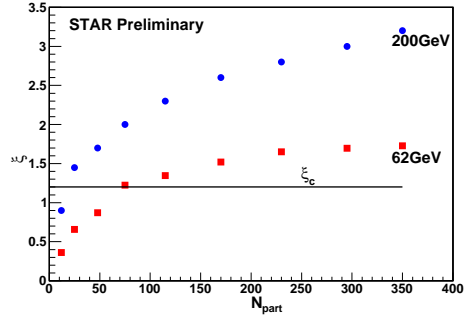}}
%\vspace*{0.0cm}
\caption{Percolation density parameter $\xi$ vs $N_{part}$} 
\label{perc3}
\end{figure}
These experimental $\xi$ values are used to get the bulk thermodynamic observables.

\section{Determination of the Temperature}

The strong longitudinal chromo-electric fields produce Schwinger-Bialas \cite{swinger1,pajares3,bialas} like radiation with a thermal spectrum, in analogy with the Hawking-Unruh radiation \cite{hawking,unruh,parikh,khar1,khar2,khar3}. Both the Schwinger-Bialas and Hawking-Unruh derivations lead to the same value of the maximum entropy temperature.
Above the critical value of $\xi$, the QGP in CSPM consists of massless constituents (gluons). 
The percolation density parameter $\xi$ determines the
cluster size distribution, the temperature T($\xi$) and the transverse momentum in the collision \cite{pajares3}.  
The connection between $\xi$ and the temperature $T(\xi)$ involves the Schwinger mechanism (SM) for particle production.  
In CSPM the Schwinger distribution for massless particles is expressed in terms of $p_{t}^{2}$
\begin{equation}
dn/d{p_{t}^{2}} \sim e^{-\pi p_{t}^{2}/x^{2}}
\end{equation}
with the average value of string tension $\langle x^{2} \rangle$. Gaussian fluctuations in the string tension around its mean value
transforms SM into the thermal distribution \cite{bialas}
\begin{equation}
dn/d{p_{t}^{2}} \sim e^{(-p_{t} \sqrt {\frac {2\pi}{\langle x^{2} \rangle}} )}
\end{equation}
with $\langle x^{2} \rangle$ = $\pi \langle p_{t}^{2} \rangle_{1}/F(\xi)$.
The temperature is given by
\begin{equation}
T(\xi) =  {\sqrt {\frac {\langle p_{t}^{2}\rangle_{1}}{ 2 F(\xi)}}}
\end{equation}
The string percolation density parameter $\xi$ which characterizes the percolation clusters also determines the temperature of the system. In this way at $\xi_{c}$=1.2 the percolation phase transition at $T(\xi_{c})$ models the thermal deconfinement transition. 
In the determination of temperature using Eq.(13) the value of $F(\xi)$ is obtained using the experimental data \cite{nucleo}. We will adopt the point of view that the experimentally determined chemical freeze-out temperature is a good measure of the phase transition temperature, $T_{c}$ \cite{braunmun}.
The single string average transverse momentum  ${\langle p_{t}^{2}\rangle_{1}}$ is calculated at $\xi_{c}$ = 1.2 with the  universal chemical freeze-out temperature of 167.7 $\pm$ 2.6 MeV \cite{bec1}. This gives $ \sqrt {\langle {p_{t}^{2}} \rangle _{1}}$  =  207.2 $\pm$ 3.3 MeV which is close to  $\simeq$200 MeV used previously in the calculation of the percolation transition temperature \cite{pajares3}.
Above $\xi_{c}$ =1.2 the size and density of the spanning cluster increases. We use the measured value of $\xi$ = 2.88 to determine the temperature, before the expansion, $T_{i}$  = 193.6$\pm$3.0 MeV of the quark gluon plasma in reasonable agreement with $T_{i}$  = 221$\pm 19^{stat} \pm 19^{sys}$ from the enhanced direct photon experiment measured by the PHENIX Collaboration\cite{phenix}.
 
\section{Bulk Thermodynamic Quantities}
Among the most important and fundamental problems in finite-temperature QCD are the calculation of the bulk properties of hot QCD matter and characterization of the nature of the QCD phase transition. 
The QGP according to CSPM is born in local thermal equilibrium  because the temperature is determined at the string level. We use CSPM coupled to hydrodynamics to calculate energy density, pressure, entropy density and sound velocity. As mentioned earlier the strings interact strongly to form clusters and produce the pressure at the early stages of the collisions, which is evident from the presence of elliptical flow in CSPM \cite{bautista}. After the initial temperature $ T > T_{c}$ the  CSPM perfect fluid may expand according to Bjorken boost invariant 1D hydrodynamics \cite{bjorken}
\begin{eqnarray}
\frac {1}{T} \frac {dT}{d\tau} = - C_{s}^{2}/\tau  \\
\frac {dT}{d\tau} = \frac {dT}{d\varepsilon} \frac {d\varepsilon}{d\tau} \\
\frac {d\varepsilon}{d\tau} = -T s/\tau \\
s =(1+C_{s}^{2})\frac{\varepsilon}{T}\\
\frac {dT}{d\varepsilon} s = C_{s}^{2} 
\end{eqnarray}
where $\varepsilon$ is the energy density, s the entropy density, $\tau$ the proper time, and $C_{s}$ the sound velocity. Above the critical temperature only massless particles are present in CSPM. 
The initial energy density $\varepsilon_{i}$ above $T_{c}$ is given by \cite{bjorken}
\begin{equation}
\varepsilon_{i}= \frac {3}{2}\frac { {\frac {dN_{c}}{dy}}\langle m_{t}\rangle}{S_{n} \tau_{pro}}
\end{equation}
To evaluate $\varepsilon_{i}$ we use the charged pion multiplicity $dN_{c}/{dy}$ at midrapidity and $S_{n}$ values from STAR for 0-10\% central Au-Au collisions with $\sqrt{s_{NN}}=$200 GeV \cite{levente}. The factor 3/2 in Eq.(19) accounts for the neutral pions.
\begin{figure}
\centering        
%\vspace*{-0.2cm}
\resizebox{0.50\textwidth}{!}{
\includegraphics{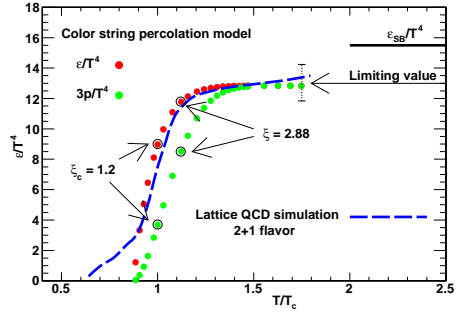}}
%\vspace*{0.0cm}
\caption{The energy density} 
\label{perc4}
\end{figure} 

The dynamics of massless particle production has been studied in QE2 quantum electrodynamics.
QE2 can be scaled from electrodynamics to quantum chromodynamics using the ratio of the coupling constants \cite{wong}. The production time $\tau_{pro}$ for a boson (gluon) is \cite{swinger}  
\begin{equation}
%\tau_{pro} = \frac {2.405\hbar}{mc^{2}}.
\tau_{pro} = \frac {2.405\hbar}{\langle m_{t}\rangle}
\end{equation}
 In CSPM the total transverse energy is proportional to $\xi$. 
From the measured value of  $\xi$ and $\varepsilon$ 
it is found  that $\varepsilon$ is proportional to $\xi$ for the range 
$1.2 < \xi < 2.88$, $\varepsilon_{i}= 0.788$ $\xi$ GeV/$fm^{3}$ \cite{nucleo,levente}. This relationship has been extrapolated to below  $\xi= 1.2$ and  above $\xi =2.88$  for the energy and entropy density calculations. Figure 3 shows $\varepsilon/T^{4}$ as obtained from CSPM along with the LQCD calculations \cite{bazavov} and the CSPM pressure $3p/T^{4}$.  

For an ideal gas of massless constituents, the energy density and pressure are related by $\varepsilon = 3P$. In LQCD the basic quantity is the interaction measure
$\Delta = \varepsilon- 3p/T^{4}$, which is also known as the trace anomaly. In CSPM
$\Delta$ reaches zero for T $ > 2 T_{c}$ while in case of LQCD it is strongly interacting even for T $ > 4 T_{c}$ \cite{bazavov}.

The sound velocity requires the evaluation of s and $ {dT}/{d\varepsilon}$, which can be expressed in terms of $\xi$ and $F(\xi)$. With $q^{1/2}$ = $F(\xi)$ one obtains:

\begin{equation}
\frac {dT}{d\varepsilon} = \frac {dT}{dq}\frac {dq}{d\xi} \frac {d\xi}{d\varepsilon}
\end{equation}
 Then $C_{s}^{2}$ becomes:
\begin{equation}
 C_{s}^{2} = (-1/4)(1+ C_{s}^{2}) \left(\frac {\xi e^{-\xi}}{1- e^{-\xi}}-1\right)
\end{equation}
 for $\xi \geq \xi_{c}$, an analytic function of $\xi$ for the equation of state of the QGP for T $\geq T_{c}$. 

\begin{figure}
\begin{minipage}[t]{0.5\textwidth}
\begin{center}
%\centering        
%\vspace*{-0.2cm}
%\resizebox{0.50\textwidth}{!}{
\includegraphics[width=7.0cm]{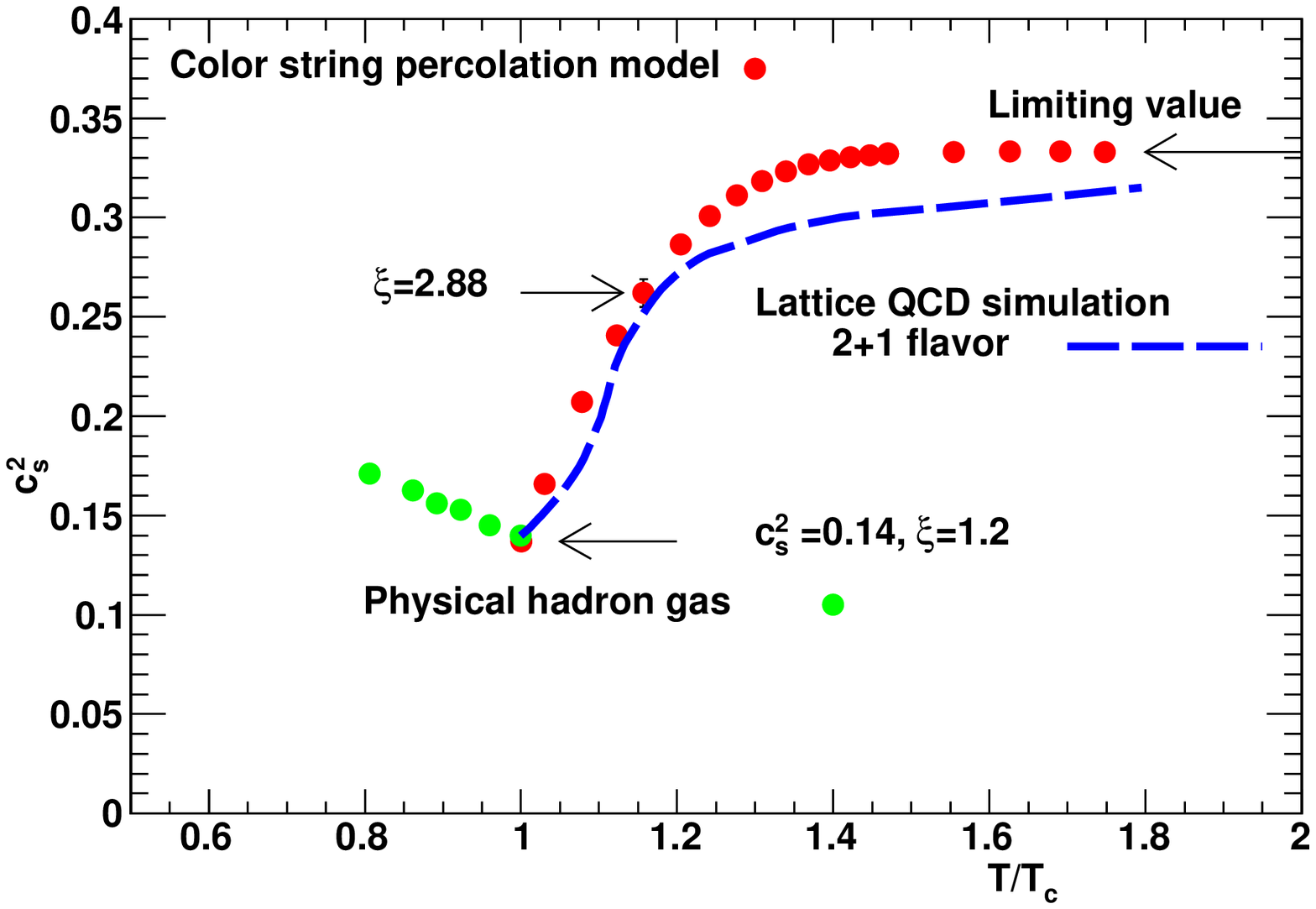}
%\vspace*{0.0cm}
%\begin{center}
\caption{The speed of sound} 
%from CSPM 
%versus $T/T_{c}^{CSPM}$(red circles) 
%and Lattice QCD-p4  speed of sound versus 
% $T/T_{c}^{LQCD}$(blue dash line)\cite{bazavov}.
 %The physical  hadron gas with resonance mass cut off  M $\leq$ 2.5 Ge
%V is shown as solid green circles \cite{satz}.} 
%\end{center}
%\label{sound velocity}
\end{center}
\hfill
\end{minipage}%
\begin{minipage}[t]{0.5\textwidth}
\begin{center}
%\centering        
%\vspace*{-0.2cm}
%\resizebox{0.50\textwidth}{!}{
\includegraphics[width=7.0cm]{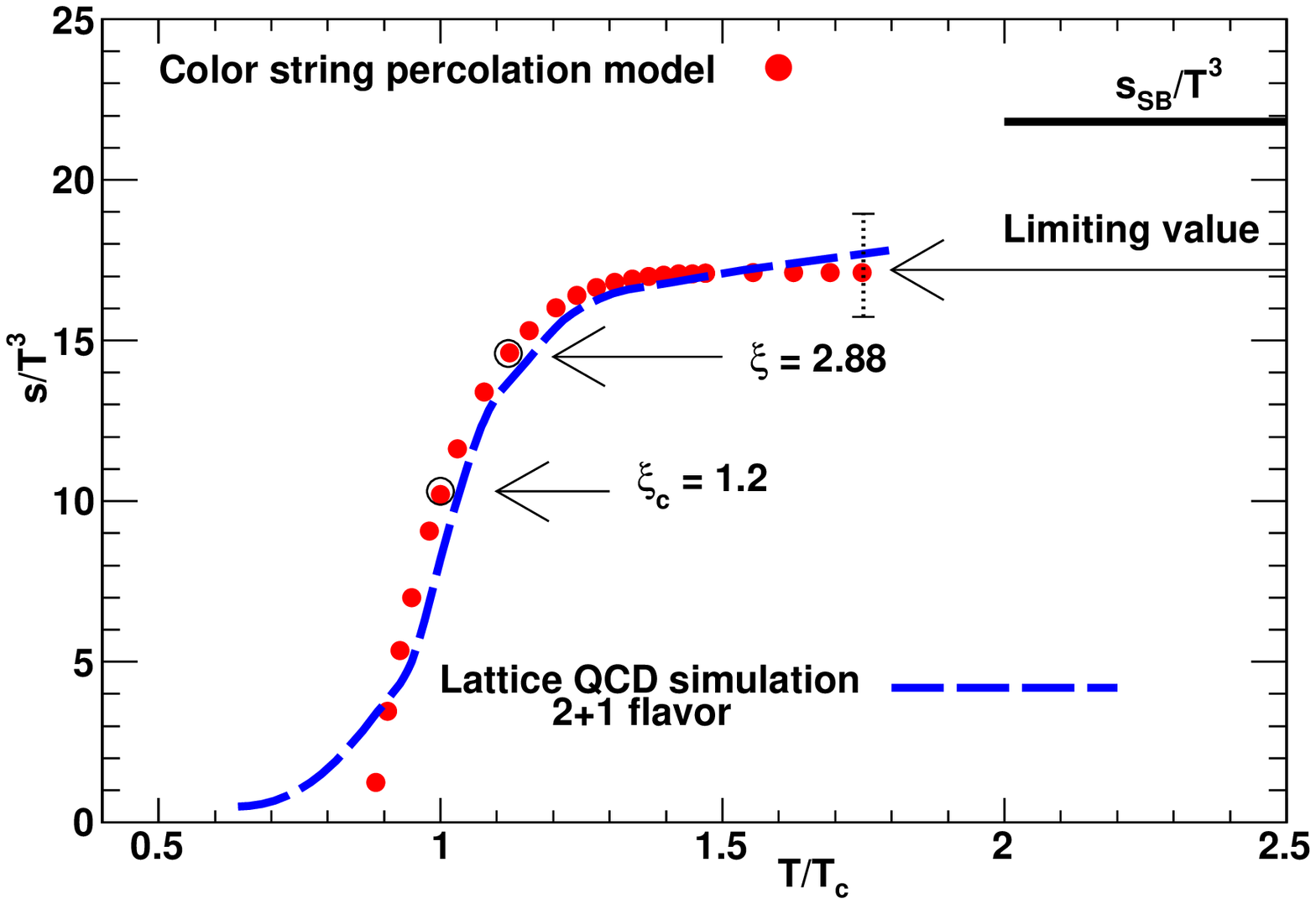}
%\hspace*{1.0cm}
%\begin{center}
\caption{The entropy density }
%from CSPM 
%versus $T/T_{c}^{CSPM}$(red circle) and Lattice QCD entropy density vs  $T/T_{
%c}^{LQCD}$(blue dash line) \cite{bazavov}.} 
%\end{center}
%\label{entropy}
\end{center}
\end{minipage}
\end{figure}

Figure 4 shows the comparison of $C_{s}^{2}$ from CSPM and LQCD. The LQCD values were obtained using the EOS of 2+1 flavor QCD at finite temperature with physical strange quark mass and almost physical light quark masses \cite{bazavov}. At $T/T_{c}$=1 the CSPM 
 and LQCD agree with the $C_{s}^{2}$ value of the physical hadron gas with resonance mass truncation M $\leq$ 2.5 GeV \cite{satz}.

The entropy density $s/T^{3}$ is obtained from energy density and speed of sound as shown in Fig.5 along with the LQCD results. CSPM is in excellent agreement with the LQCD calculations in the phase transition region for $T/T_{c} \leq $1.5. 

\section{Summary}
The percolation analysis of the color sources applied to STAR data at RHIC provides a compelling argument that the QGP is formed in central Au-Au collisions at $\sqrt{s_{NN}}=$ 200 GeV. It also suggests that the QGP is produced in all soft high energy high multiplicity collisions when the string density exceeds the percolation transition.
The results are also in agreement with lattice QCD in the phase transition region, when the results are plotted with respect to $T/T_{c}^{CSPM}$ and  $T/T_{c}^{LQCD}$. The value of $C_{s}^{2}$=0.14 is in agreement with the physical resonance gas value at the critical temperature. Thus clustering and percolation can provide a conceptual basis for the QCD phase diagram which is more general than symmetry breaking \cite{satzx}.
 
\section{Acknowledgement}
This research was supported by the Office of Nuclear Physics within the U.S. Department of Energy  Office of Science under Grant No. DE-FG02-88ER40412.

\end{document}